\documentclass[onecolumn]{gji}
\let\bibhang\relax
\usepackage{natbib}
\usepackage{timet}
\usepackage{amsmath}

\usepackage{graphicx}
\usepackage{amssymb}
\usepackage{color}
\newcommand{\bx}{{\bf x}}

\newcommand{\bm}{{\bf m}}

\newcommand{\ener}{{\varepsilon}}

\newcommand{\bnabla}{{\boldsymbol{\nabla}}}
\newcommand{\bmu}{{\boldsymbol{\mu}}}
\newcommand{\blamb}{{\boldsymbol{\Lambda}}}

\newcommand{\cc}{{\mathcal C}}
\newcommand{\ccref}{{\mathcal C}_{\rm ref}}

\newcommand{\pspec}{{\mathcal P}}

\title[Measurements and Kernels for Source-Structure Inversions in Noise Tomography]
  {Measurements and Kernels for Source-Structure Inversions in Noise Tomography}
\author[Shravan M. Hanasoge]
  {Shravan M. Hanasoge\thanks{hanasoge@princeton.edu} \\
  Department of Geosciences, Princeton University, NJ 08544, USA\\
  Max-Planck-Institut f\"{u}r Sonnensystemforschung, 37191 Katlenburg-Lindau, Germany\\
  Tata Institute of Fundamental Research, Mumbai 400005, India
  }
\date{Received 2013 June 18; in original form 2013 April 12}

\begin{document}

\label{firstpage}

\maketitle

\begin{summary}
Seismic noise cross correlations are used to image crustal structure and heterogeneity. 
Typically, seismic networks are only anisotropically illuminated by seismic noise, a consequence of the non-uniform distribution
of sources. Here, we study the sensitivity of such a seismic network 
to structural heterogeneity in a 2-D setting. We compute finite-frequency cross-correlation sensitivity kernels for travel-time, waveform-energy 
and waveform-difference measurements. In line with expectation, wavespeed anomalies are best imaged using travel times and the source distribution
using cross-correlation energies. Perturbations in attenuation and impedance are very difficult to image and reliable inferences require a high degree
of certainty in the knowledge of the source distribution and wavespeed model (at least in the case of transmission tomography studied here).
We perform single-step Gauss-Newton inversions for the source distribution and the wavespeed, in that order, and quantify the associated
Cram\'{e}r-Rao lower bound. The inversion and uncertainty estimate are robust to errors in the source model but are sensitive to the theory used to interpret of measurements. 
We find that when classical source-receiver kernels are used instead of cross-correlation kernels, errors appear in the
both the inversion and uncertainty estimate, systematically biasing the results. We outline a computationally tractable algorithm to account for distant
sources when performing inversions.
\end{summary}

\section{Introduction}
Terrestrial seismic noise fluctuations are observed over a broad range of temporal frequencies, excited by
anthropogenic activity,
storms and oceanic microseisms among other mechanisms \citep[e.g.,][]{longuet50, nawa98, kedar05, rhie04, stehly06}.
Seismic noise has been used as a successful alternative to earthquake tomography to image the crust. 
It has shown great promise as a means of generating new crustal constraints 
and characterizing its temporal variations \citep[e.g.,][]{campillo07, wegler07,BrenguierSCI08,shapiro11, rivet11}.

The dominant measurement in seismic noise tomography is the ensemble-averaged cross correlation.
The cross correlation measurement, by virtue of being a product of wavefields, possesses different physical 
attributes when compared with its classical tomographic analog, i.e., wavefield displacement. 
Under certain ideal circumstances,
such as when the source distribution is uniform, or when wave attenuation and source generation are
collocated, it is possible to derive representation theorems \citep[e.g.,][]{snieder10} whichs state that the cross correlation
is effectively Green's function between the stations. Based on this reasoning, practitioners in this field view cross correlation
measurements through the lens of classical tomography.
 Earth noise sources are however typically highly
anisotropically distributed, as a consequence of which these theorems are broken and a more rigorous interpretation becomes necessary. 

The theoretical treatment of terrestrial seismic noise has connections with the wavefields
of stars, specifically, the Sun. The use of cross correlations of the wavefield of the Sun to probe the interior 3-D structure of the Sun
was pioneered by \citet{duvall}.
\citet{gizon02}, based on the finite-frequency theory of \citet{dahlen99},
derived a recipe to compute kernels for cross correlations of helioseismic noise arising from a distribution of sources. 
Building on this work, \citet{tromp10} derived an adjoint theory for cross correlations for the terrestrial case. 
The theory allows for treating distributions of 
sources as opposed to a discrete number of them \citep[e.g.,][]{larose06, tsai09}. The results of \citet{tromp10} enable the
prediction of cross correlations on a rigorous basis, for a given Earth model and source distribution 
\citep[this problem has been studied extensively by, e.g.,][]{hellegji06,ChevrotJGR07,YangG308,WeaverJASA09,cupillard10,tsai10,FromentGEO10}.

{However, complex random heterogeneities in Earth's crust induce multiple scattering \citep[e.g.,][]{wegler07} introduce 
theoretical complications in the forward calculation. 
In a scenario where the randomness in the medium is ergodic, i.e., sampling some finite area of the medium
is sufficient to extract its full properties, and when the probability density function 
describing the heterogeneities is spatially translationally invariant, a variety of methods can be brought to bear on
the forward problem, including periodic homogenization \citep[e.g.,][]{varadhan82} and radiative transfer \citep[e.g.,][]{sato12}.
Modeling the heterogeneous Earth, whose probability density function is likely to vary from one point to another, is considerably more difficult \citep[e.g.,][]{capdeville10, sato12}.}

{The first Born approximation treats a scatterer as a source of a new wave (where the type and amplitude
of the new source, i.e., monopolar, dipolar etc. depends on the type of scatterer) and therefore, roughly speaking, medium randomness tends to `isotropize' the directionality of the sources.
Consequently, the impact of structural heterogeneities on the source distribution can certainly be modeled when inverting for the azimuthal distribution of source amplitudes. 
Structural heterogeneities outside of the aperture of the network are unlikely to affect the local measurement process and may, for all practical purposes, be lumped
into an effective source term.}

The purpose of this article is to investigate the robustness of noise-tomographic inferences to incomplete theoretical models (such as treating noise
as if it were a classical tomographic measurement) and ignoring the anisotropy of illumination. 
We do so using the framework of a 2-D wave propagation problem with heterogeneities such as wavespeed and attenuation illuminated by anisotropic sources.
The governing equation is outlined in section~\ref{gov.eq}, followed by a discussion of the cross correlation (section~\ref{computecc}),
choice of measurements (section~\ref{measure.choice}) and a means to compute sensitivity kernels for the parameters of density, attenuation,
wavespeed and source distribution for a given measurement. We calculate the response of a network of stations to anomalies in density, wavespeed and attenuation for 
uniform and non-uniform source distributions and demonstrate that the ability to image attenuation and density is sensitively dependent on the knowledge
of the sources. Using cross-correlation energy and travel times as measurements, we pose inverse problems for the source distribution and wavespeed, 
solving them in that order. We invert for these parameters using a 
single-step Gauss-Newton method. The appendices outline the procedure applied to compute kernels.
Synthetics around the heterogeneous model show that the misfit falls by a factor of more than 4. 
Finally, we quantify the uncertainty of the inversion by computing the diagonal of Hessian matrix, to which we have full access.

\section{The 2-D forward problem}\label{gov.eq}
We consider wave propagation in a 2-D heterogeneous medium described by
\begin{equation}
\rho\partial_t^2\phi  + \rho\Gamma\,\partial_t\phi -  \bnabla\cdot(\rho\,c^2\bnabla\phi) = {f(\bx,t)},\label{phi.eq}
\end{equation}
where $\phi(\bx,t)$ is the wavefield, $\rho(\bx)$ the density of the medium, $c(\bx)$ the wavespeed, $\Gamma(\bx)$ the attenuation (measured in
units of inverse time), $\bnabla$ the covariant
spatial derivative, $t$ time, $\partial_t$ the derivative with respect to time, ${f}$ the source and $\bx$ the spatial coordinate.
The form of attenuation in equation~(\ref{phi.eq}) is simplistic, does not fully capture the mechanisms
that govern wave damping in Earth, and is only adopted for the ease it affords in studying the sensitivity of noise to attenuation. 
See, e.g., \citet{zhu_attenuation}, for details on a more rigorous treatment of wave attenuation.
Equation~(\ref{phi.eq}) is solved using a pseudo-spectral scheme. Fast Fourier transforms are applied to compute spatial derivatives and
temporal evolution of the system is achieved using an optimized second-order Runge-Kutta scheme \citep{hu}. Green's functions are
computed around each station and cross correlations are estimated according to the method described by, e.g., \cite{tromp10}, \cite{hanasoge12_sources}.
We discuss the method here for the sake of completeness in section~\ref{computecc}. See equation~(\ref{cc.theory}) for the general relationship between Green's functions
and the cross correlation.

\subsection{Computing Noise Cross correlations}\label{computecc}
The first step in the inverse problem for noise is to have a means of predicting cross correlations.
The formal interpretation of noise cross correlations in terms of Green's functions of the medium for an arbitrary distribution of sources has been 
described by, e.g., \citet{woodard}, \citet{gizon02}, \citet{hanasoge11} for helioseismology and by, e.g., \citet{tromp10}, \citet{hanasoge12_sources} for noise tomography.
We summarize the theoretical results here since they will be used frequently, referring the interested reader to these references for more detailed expositions. 
For a temporally stochastic and spatially stationary and uncorrelated distribution of sources, a situation relevant in some cases to Earth ambient noise, it may be shown that the
expected cross correlation is connected to Green's functions of the medium thus (invoking identity~[\ref{cross.c}])
\begin{equation}
\cc(\bx_i,\bx_j, \omega) = \int d\bx'\,G^*(\bx_i,\bx'; \omega)\,G(\bx_j,\bx'; \omega)\, S(\bx')\,\pspec(\omega),\label{cc.theory}
\end{equation}
where $\pspec(\omega)$ is the temporal power spectral distribution of the sources, $S(\bx')$ the spatial distribution of sources,
$G(\bx, \bx',\omega)$ is Green's function measured at point $\bx$ due to a source at $\bx'$. The quantity $\cc$ is the cross correlation
of signals measured at receivers $\bx_i, \bx_j$. Note that $\cc$ is the expected value of the cross correlation, and that assumptions
of stationarity and ergodicity have been invoked in justifying its existence \cite[e.g.,][]{hanasoge12_sources}. Equation~(\ref{cc.theory})
is written in frequency domain for the sake of simplicity. The corresponding form in time domain is more tedious and we will avoid it in
this article.

Source-receiver seismic reciprocity gives us,
\begin{equation}
G(\bx, \bx'; \omega) = G(\bx', \bx; \omega),
\end{equation}
since we have an unbounded medium. This result allows us to rewrite equation~(\ref{cc.theory}) thus
\begin{equation}
\cc(\bx_i,\bx_j, \omega) = \int d\bx'\,G^*(\bx',\bx_i; \omega)\,G(\bx',\bx_j; \omega)\, S(\bx')\,\pspec(\omega),
\end{equation}
an important manipulation that greatly simplifies the computation 
since all we need to estimate the predicted cross correlation is to calculate Green's functions due to 
sources placed at the receivers and integrate the product weighted by the source distribution.

For a uniform background medium, which we use as the starting model, the following operative relation holds
\begin{equation}
G(\bx,\bx'; \omega) = G(|\bx - \bx'|; \omega). \label{invar.tran}
\end{equation}
This relationship makes it computationally feasible to compute spatial convolutions of Green's functions by
transforming to the Fourier domain. Note that equation~(\ref{invar.tran}) does not hold for heterogeneous models.

\subsection{Measurements}\label{measure.choice}
The dominantly used measurement in noise tomography is the ensemble-averaged cross correlation, defined by
\begin{equation}
\cc_{ij}(t) = \frac{1}{T}\int_0^T\,dt'\,\phi(\bx_i,t+t')\,\phi(\bx_j,t'),\label{cc.def}
\end{equation}
where $\cc_{ij}(t)$ is the cross correlation at time lag $t$ between signals $\phi$ recorded at stations located at $\bx_i$ and $\bx_j$ using a 
recording length $T$. To improve the signal-to-noise ratio of the measurement, an ensemble average over several temporal epochs of length $T$
is taken. Processing methods such as pre-whitening \cite[e.g.,][]{seats12} and one-bit filtering \cite[e.g.,][]{aki57} are typically used to make sensible measurements.
From these measurements, wave travel times, amplitudes, and cross correlation energies may be extracted. Here, we study the impact of this 
standard suite of measurements, i.e., travel times, cross correlation energies, and waveforms, on the inversions.
The travel-time shift $\tau$, formally defined in Appendix~\ref{varmisfit}, has been studied by a number of authors \cite[e.g.,][]{luo91, woodard, dahlen99, gizon02, tromp05}.
\begin{figure}
\centering
\includegraphics*[width=\linewidth]{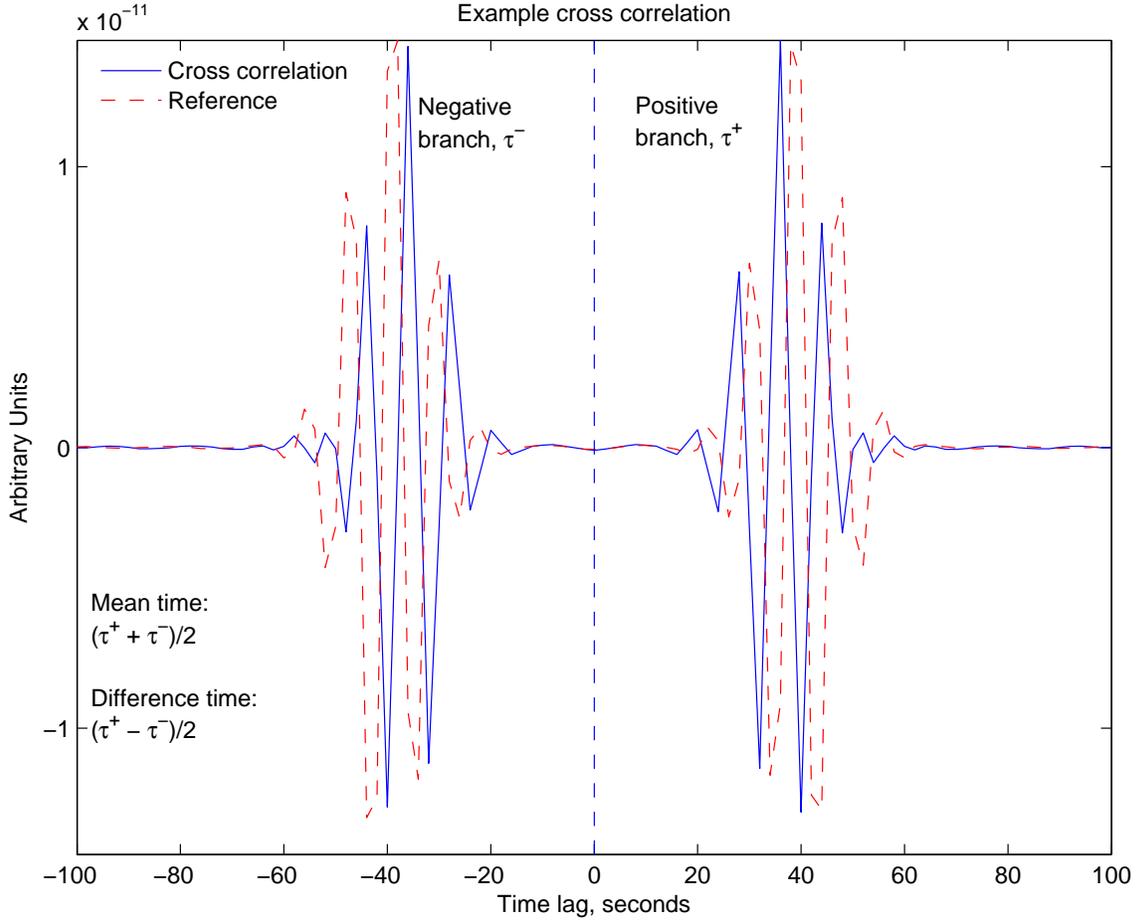}
\caption{Example cross correlation and the reference measured at a pair of stations separated by a distance of 120 km, where the wavespeed of the medium is 3 km/s.
The stations are placed symmetrically across a disc-shaped wavespeed anomaly of +10\%, with no density
or attenuation contrasts.
 The horizontal axis is time lag measured in seconds. The positive and negative branches correspond to positive and negative time lags respectively.
 Travel times measured in each branch are denoted by $\tau^+$ and $\tau^-$. The average and difference of these two travel times are termed
 mean and difference times respectively (Eq.~[\ref{mean.diff}]). \label{exam.cc}}
\end{figure}
The cross correlation function has two branches, the positive-lag branch ($t>0$) and the negative-lag branch ($t<0$). Thus two
travel times may be extracted from each cross correlation, and we denote $\tau^+$ as the travel time obtained from the positive
branch and $\tau^-$ as the travel time associated with the negative branch.
In the inversion, we will use {\it mean and difference travel times}, which are defined as \cite[e.g.,][]{gizon02, tromp10}
\begin{equation}
\tau^m = \frac{\tau^+ + \tau^-}{2},~~~~~~~~~~~\tau^d = \frac{\tau^+ - \tau^-}{2}.\label{mean.diff}
\end{equation}
The energy of the cross correlation is given by \cite[e.g.,][]{dahlen02, hanasoge12_sources}
\begin{equation}
\ener = \sqrt{\frac{1}{T}\int dt\,w(t)\,\cc^2},
\end{equation}
and the waveform difference is $w(t)(\ccref-\cc)$. Thus we have four measurements from a cross correlation:
{\it mean travel time, difference travel time, energy}, and {\it waveform difference}. In order, we define
misfit functionals for mean travel-time shifts, difference travel-time shifts, energy and waveform 
\begin{eqnarray}
\chi^m &=& \frac{1}{2N}\sum_{ij} (\tau^{m}_{ij} )^2,\label{mean.tt}\\
\chi^d &=& \frac{1}{2N}\sum_{ij} (\tau^{d}_{ij} )^2,\\
\chi^\ener &=& \frac{1}{2N}\sum_{ij} \left[\ln\left(\frac{\ener^{\rm ref}_{ij}}{\ener_{ij}}\right)\right]^2,\label{misfit.ener}\\
\chi^{\rm w} &=& \frac{1}{2N}\sum_{ij} \int dt\, w(t)\,(\ccref-\cc )^2,
\end{eqnarray}
where $\ener^{\rm ref}$ is the energy of the reference cross correlation and $N$ is the number of measurements. Studying variations of these misfit functionals allows us 
to evolve our starting model; pertinent details are discussed in Appendices~\ref{varmisfit} and~\ref{kern.comp}.
In this problem, we use cross correlations computed using equation~(\ref{cc.theory}) for heterogeneous media
as `data' inputs. As a starting model for the inversion, we use a uniform homogeneous model with density, wavespeed and attenuation given by $\rho_e, c_e, \Gamma_e$.
The problem is to obtain estimates for $\rho_d, c_d, \Gamma_d$. 

\section{Kernels}\label{compute.kernel}
\begin{figure}
\centering
\includegraphics*[width=\linewidth]{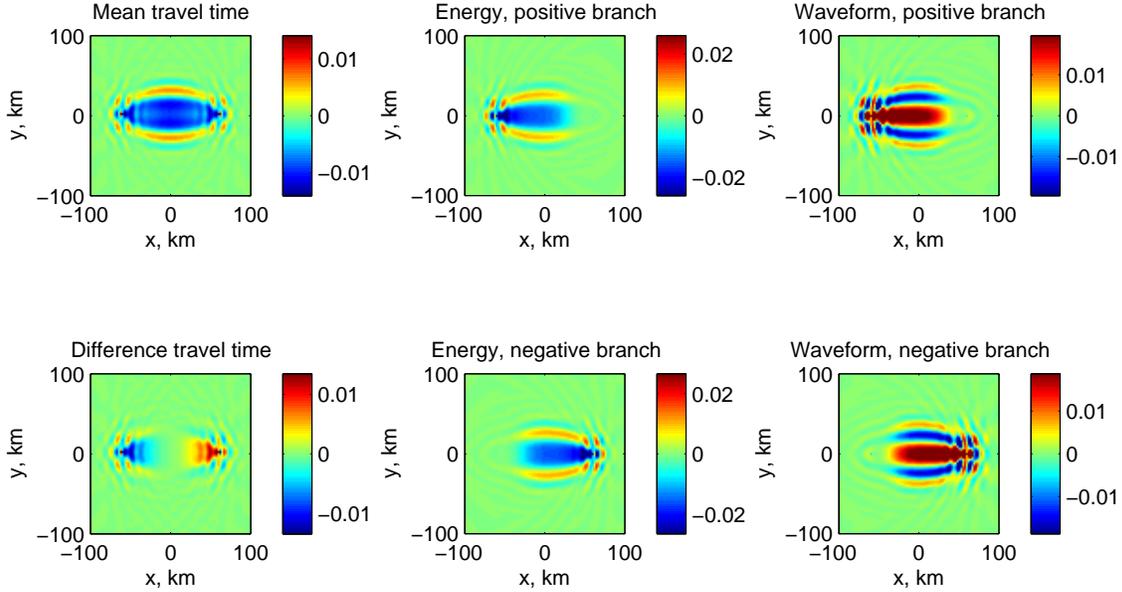}
\caption{Example wavespeed kernels for energy,  waveform difference, mean and difference travel-time measurements for a station pair located 120 km apart.
The units of the travel-time kernels are ${\rm s\, km}^{-2}$ and the energy kernels are ${{\rm km}^{-2}}$ (since we are dealing with a 2-D problem).
The integral of the difference travel time kernel is zero and is not sensitive to wavespeed anomalies. The energy kernel is interesting in that it does not have a `doughnut hole'
in the center, which the travel-time kernels evidently possess {\cite[as][have noted]{ dahlen02, nolet05}}.
Waveform differences are also sensitive to the underlying anomaly, also appearing to not possess a `hole'. 
The source distribution is uniform in this case, resulting in strong symmetries about the bisector between and line joining the two stations.
\label{exam.kernels}}
\end{figure}

\begin{figure}
\centering
\includegraphics*[width=\linewidth]{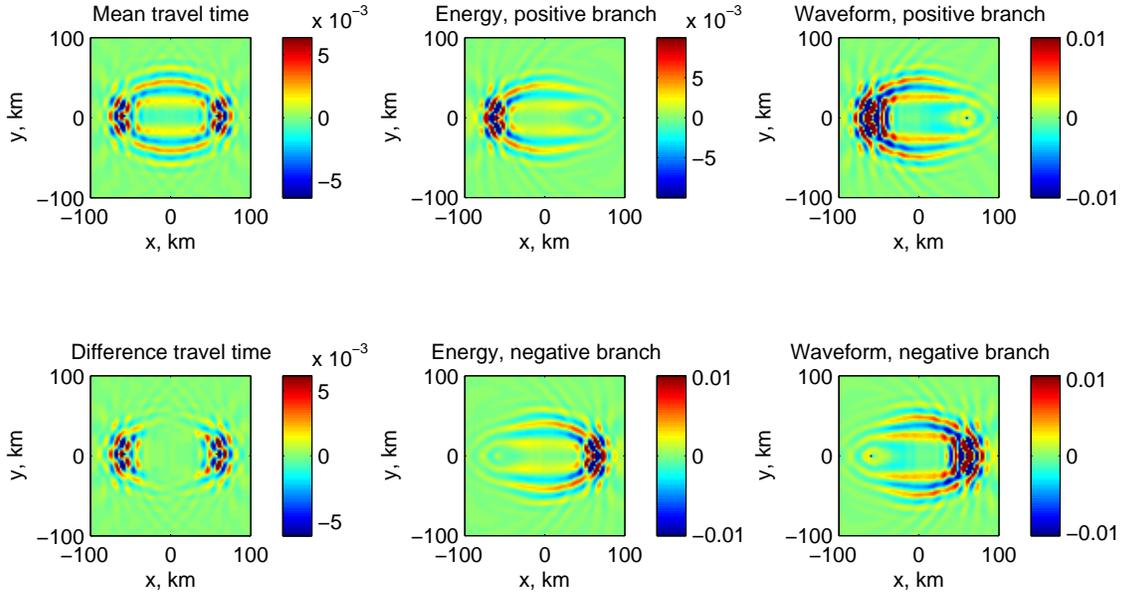}
\caption{Example impedance kernels for energy,  waveform difference, mean and difference travel-time measurements for a station pair located 120 km apart. 
The units of the travel-time kernels are ${\rm s\, km}^{-2}$ and the energy kernels are ${{\rm km}^{-2}}$ (since we are dealing with a 2-D problem). Not surprisingly, 
it is seen from the color bars that the impedance kernels do not contribute as significantly to the misfit as the wavespeed kernels. The source distribution is uniform in this case, resulting in strong symmetries about the bisector between and line joining the two stations.
\label{exam.kernels2}}
\end{figure}

With choices for measurements and a starting model, we are ready to compute finite-frequency kernels. A general computational adjoint theory for noise kernels
was described by \citet{tromp10}; here, we consider a simpler problem, where the starting model is homogeneous. Let us
consider the variation of the mean travel-time misfit functional~(\ref{mean.tt})
\begin{equation}
\delta\chi^m = \frac{1}{N}\sum_{ij} \tau^{m}_{ij}\delta\tau^{m}_{ij} = \frac{1}{2N}\sum_{ij} \tau^{m}_{ij}(\delta\tau^{+}_{ij} + \delta\tau^{-}_{ij}).
\end{equation}
We analyze the variation of the positive travel-time shift keeping in mind that a similar procedure is applied to the negative time shift. Substituting
the formal definition of the travel-time from equation~(\ref{tt.def}), dropping the `+' superscript for convenience and invoking identity~(\ref{time.integral}),
\begin{equation}
\delta\tau = \frac{\int dt\,w(t)\,{\dot\ccref}\,\delta\cc(t)}{\int dt\,w(t){\dot\ccref}^2} = {\int d\omega\, \mathcal{W}^*(\omega)\,\delta\cc(\omega)},\label{tau.del}
\end{equation}
where the windowing function, reference cross correlation and the denominator have been subsumed into a weight function, whose
temporal Fourier transform is written as  ${\mathcal W}(\omega)$ . The transformation to Fourier domain was accomplished according to the 
Fourier convention described in Appendix~\ref{convention}. From this point, we only broadly sketch the details of how to compute the kernel.
Details of the derivation are found in, e.g., \citet{hanasoge12_sources} and in Appendices~\ref{varmisfit} and~\ref{kern.comp}. 
The variation of the cross correlation function is further expressed in terms of variations of Green's functions i.e., misfit arising from imperfect 
knowledge of structure and the source distribution,
\begin{eqnarray}
\delta\cc &=& \delta\cc_{\rm structure} + \delta\cc_{\rm source},\\
\delta\cc_{\rm structure} &=& \int d\bx' [\delta G^*(\bx_i, \bx'; \omega)\, G(\bx_j, \bx'; \omega) + G^*(\bx_i, \bx'; \omega)\, \delta G(\bx_j, \bx'; \omega)]\,S(\bx')\pspec,\label{struc.del}\\
\delta\cc_{\rm source} &=& \int d\bx' G^*(\bx_i, \bx'; \omega)\, G(\bx_j, \bx'; \omega) \delta S(\bx') \,\pspec.
\end{eqnarray}
The variation of Green's function as described by the single-scattering first-Born approximation is
\begin{equation}
\delta G(\bx_j,\bx';\omega) = -\int d\bx{''} G(\bx_j,\bx{''})\, (\delta{\mathcal L}\, G)(\bx{''},\bx{'}),
\end{equation}
where $\delta{\mathcal L}$ is the variation of the wave operator. Some algebraic manipulation is required to derive expressions for the kernels; see appendix~\ref{kern.comp}
for a more detailed discussion on this. The kernels are designed to address non-dimensional quantities and are connected to the variation of the misfit (for a given measurement) thus
\begin{equation}
\delta\chi = \int d\bx \,K_c(\bx)\,{\delta\ln c} + K'_\rho(\bx)\,{\delta\ln \rho} + K_\Gamma(\bx)\,{\delta\ln \Gamma} + K_S\,\delta S,
\end{equation}
where $K_c,K'_\rho, K_\Gamma, K_S$ are the wavespeed, impedance, attenuation and source kernels respectively (see Appendix~\ref{kern.comp} for
full expressions). In Figure~\ref{exam.kernels}, we show examples of
wavespeed kernels for all the measurements discussion in section~\ref{measure.choice}.
Similarly in Figure~\ref{exam.kernels2} we show impedance kernels for these measurements.

\section{The total misfit gradient}\label{expts1}
\begin{figure}
\centering
\includegraphics*[width=\linewidth]{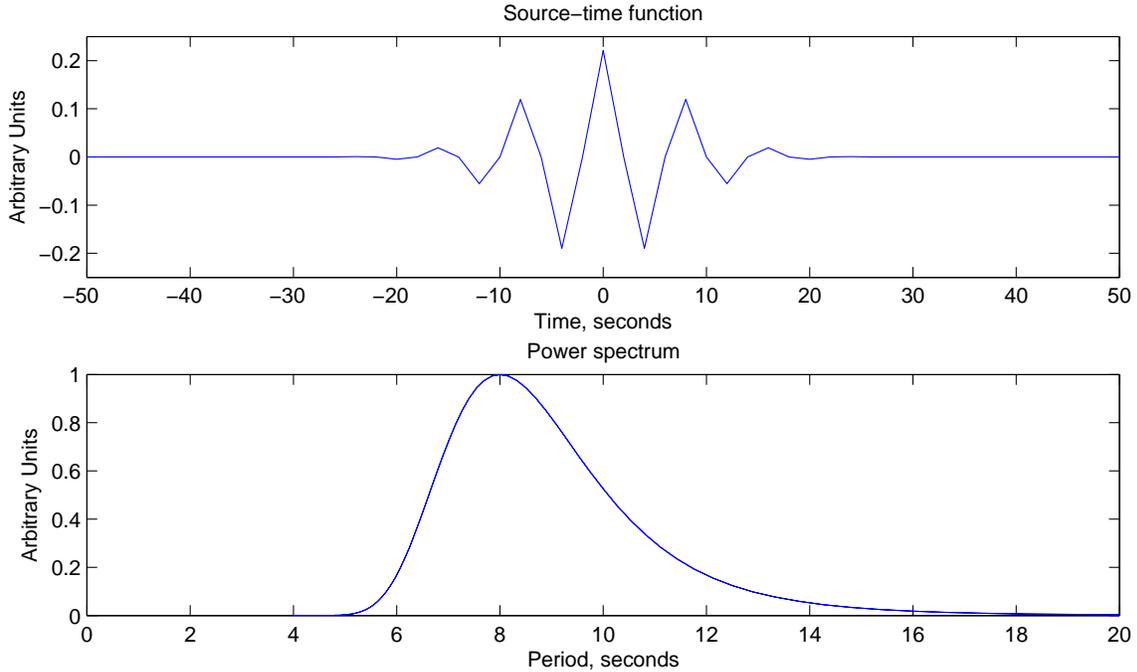}
\caption{Source-time function and power spectrum for all the kernels shown and the inversions.
In this problem we focus only on this specific frequency band.
\label{spec}}
\end{figure}

Kernels, measurements and `data' (numerically computed around the heterogeneous model) in hand, we are ready to study the inverse problem. 
For all the problems, the starting
structure model is a homogeneous uniform medium and the source model consists of a uniform ring.
In the first pass at an inverse problem, we compute the total misfit gradient, which is an indicator
of how well the inversion is likely to work.
We explore three variants here:
\begin{enumerate}
\item Only wavespeed perturbation, true source distribution is uniform, starting source distribution model = true source distribution,
\item Only attenuation perturbation (+20\%) within the central region, true source distribution is uniform, starting source distribution model = true source distribution.
\item Only wavespeed perturbation, true source distribution is non-uniform, starting source distribution model = uniform,
\end{enumerate}
The specific frequency band of 6-12 seconds is
considered here. The source-time function and the power spectrum are shown in Figure~\ref{spec}.
{Note that, when considering dispersive surface waves on Earth, appropriate filters may
be applied to narrow the frequency range, thereby allowing for defining the measurements with greater clarity. The theory discussed here directly applies to frequency-filtered
measurements, only requiring the filter to be incorporated in the definition of the cross correlation \citep[see, e.g.,][for details]{tromp10}.}

\subsection{Uniform distribution of sources}
\begin{figure}
\centering
\includegraphics*[width=\linewidth]{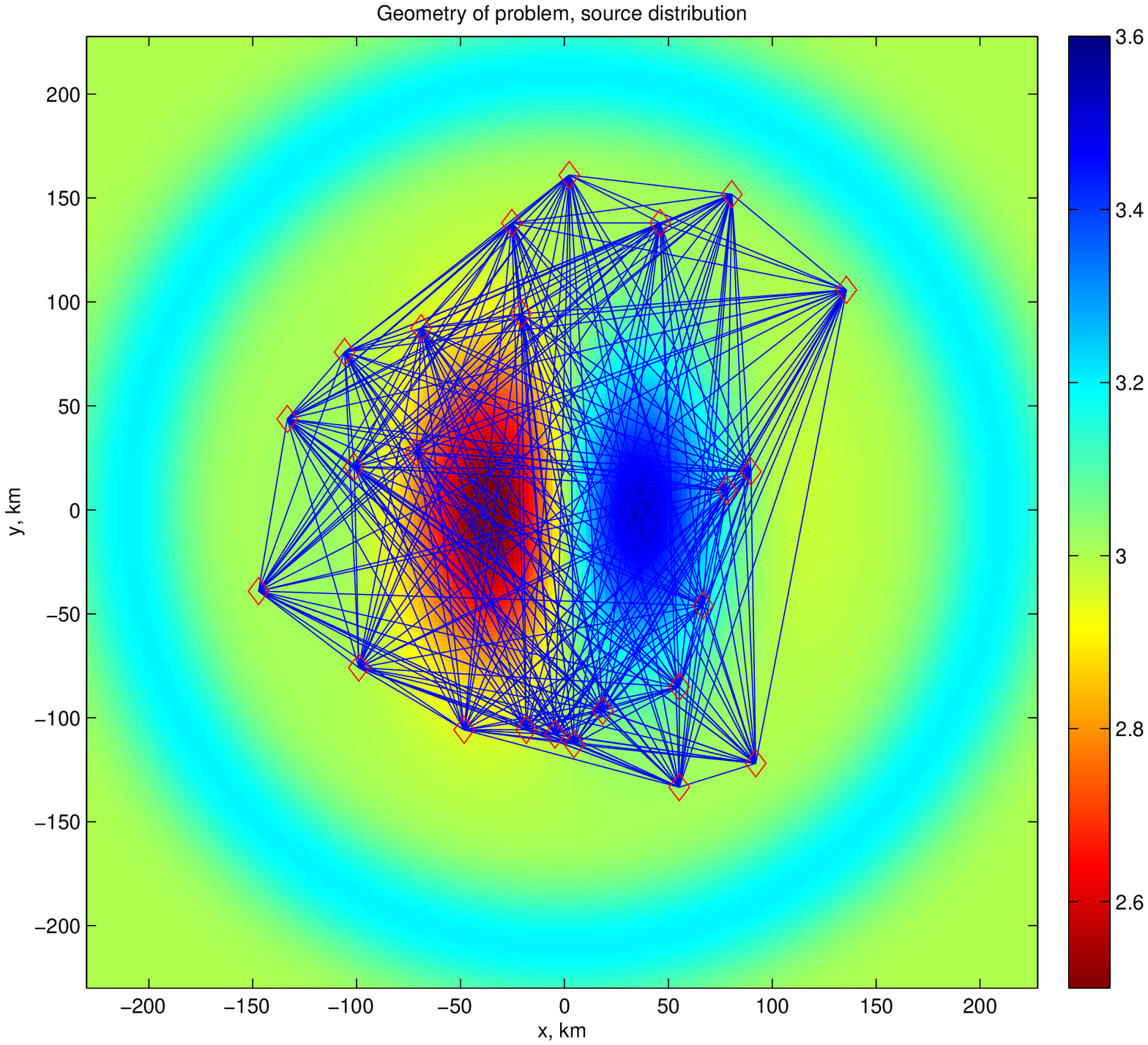}
\caption{The starting model consists of a ring-like uniform distribution sources (light blue ring) surrounding the network.
The ray-path coverage of the network is excellent within the central wavespeed anomaly, which
consists of positive and negative perturbations to the tune of 20\%, suggesting that the eventual inversion will likely be good. Stations, of which there are 24, are marked by diamonds.
Note that despite the uniform illumination, the ring of sources is too close to the network for representation theorems
to accurately apply \citep[e.g.,][]{snieder10}.
\label{source_uniform}}
\end{figure}

The first case we consider
is one in which there is a $\pm 20\%$ wavespeed perturbation within the central region while the density is uniform and the 
attenuation level is fixed at a quality factor of $Q = 100$. 
Figure~\ref{source_uniform} shows a uniform distribution of sources surrounding a network of 24 stations
which in turn surround the anomaly. This scenario is ideal because of the illumination and coverage.
We emphasize that despite the uniform illumination, the ring of sources is too close to the network for representation theorems, 
which allow for the cross correlation to be written as a linear combination of Green's functions, to accurately apply \citep[e.g.,][]{snieder10}.

\begin{figure}
\centering
\includegraphics*[width=\linewidth]{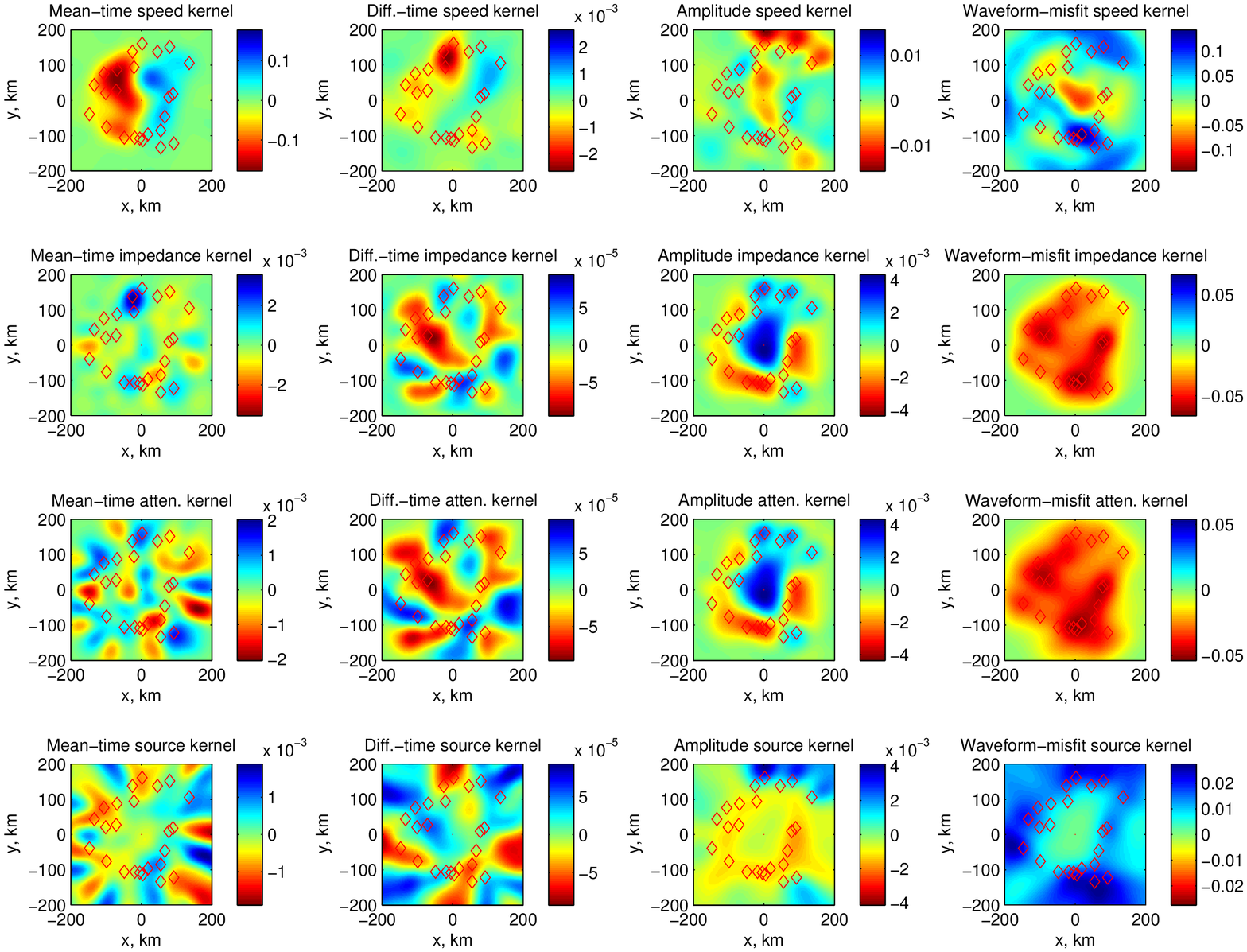}
\caption{Total gradients of misfit functions comprising different measurements with respect to model parameters. The true anomaly is in wavespeed.
This is a best-case scenario of sorts because we assume that the true source distribution is the same as in Figure~\ref{source_uniform}.
The sum of all event kernels, in some ways, is the inversion, since it is the gradient of the entire misfit functional.  A 3-point Gaussian smoothing filter was applied to the kernels.
There are four
parameters that we have imperfect knowledge of and that we wish to invert for: wavespeed, impedance, attenuation,
and source distribution (rows, top to bottom). We have four different measurements, mean and difference travel times, energies and waveform misfit (columns, left to right).
The true model contains positive and negative wavespeed perturbations to the tune of $\pm 20\%$ within the central region. It is seen that the mean travel-time is able to infer the wavespeed. That said, waveform differences also indicate attenuation and impedance
anomalies, both of which are non-existent. The different magnitudes of the total gradient generally indicate the relative significance
of each kernel (with the exception of waveform difference which has not been normalized).
\label{event_wavespeed}}
\end{figure}

In Figure~\ref{event_wavespeed} we show the misfit gradients of the four measurements (mean
and difference travel times, amplitudes and waveform difference) with respect to four model
parameters (wavespeed, impedance, attenuation and source distribution) for case (i), illustrated in Figure~\ref{source_uniform}. 
Mean travel times successfully point to the anomaly and waveform differences are problematic, pointing
misleadingly to an attenuation anomaly. A 3-point Gaussian smoothing filter was applied to the kernels.

\begin{figure}
\centering
\includegraphics*[width=\linewidth]{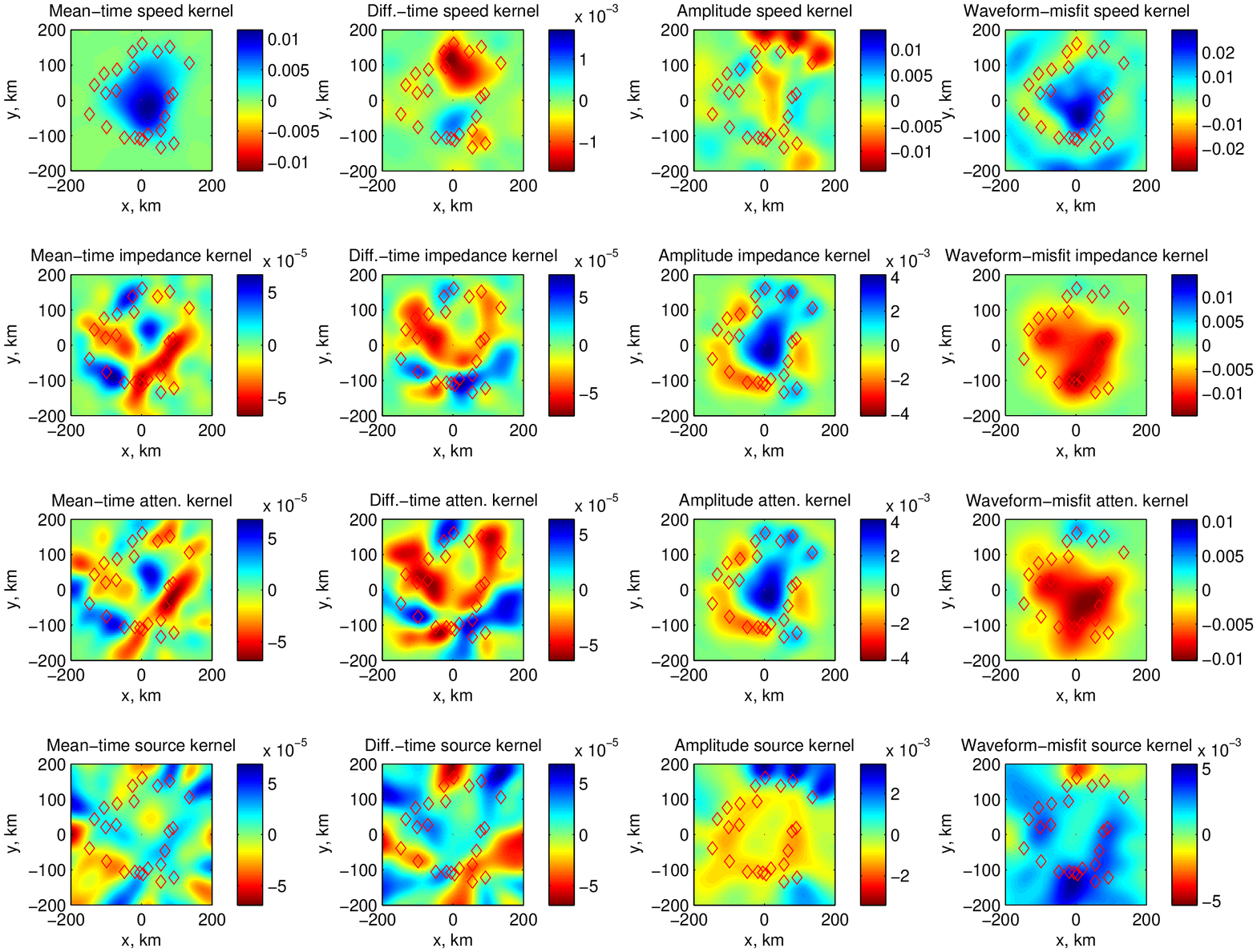}
\caption{Total gradients of misfit functions comprising different measurements with respect to model parameters. The true anomaly is in attenuation.
This is a best-case scenario of sorts because the true source distribution is as in Figure~\ref{source_uniform}.
There are four
parameters that we have imperfect knowledge of and that we wish to invert for: wavespeed, impedance, attenuation,
and source distribution (rows, top to bottom). We have four different measurements, mean and difference travel times, energies and waveform misfit (columns, left to right).
The true anomaly is a 20\% increase in attenuation in the central region of Figure~\ref{source_uniform}, with excellent ray coverage. And yet, looking at the total gradients
suggests the inversion will not meet with the success evidenced in Figure~\ref{event_wavespeed}.
The mean travel times indicate a wavespeed anomaly rather than in attenuation. 
Amplitude measurements are somewhat more successful in pointing out an attenuation anomaly although there appears to be
a tradeoff with density. Attenuation and density, as is well known, are indeed difficult parameters to infer because for propagating waves,
wavespeed is overwhelmingly influential. A 3-point Gaussian smoothing filter was applied to the kernels.
\label{correct_sources}}
\end{figure}

The next case addresses the imaging of a +20\% attenuation perturbation within the central region of Figure~\ref{source_uniform}, i.e., case (ii). The assumed
and true source distributions are identical, as shown in Figure~\ref{source_uniform}. Even in this ideal scenario, Figure~\ref{correct_sources}
highlights a mixed outcome, with travel-time measurements indicating an increase in wavespeed whereas amplitude measurements
appear to implicate both density and attenuation anomalies, suggesting a tradeoff between the two. Attenuation and density are indeed
difficult to convincingly image, even with perfect source and wavespeed models \citep[see also, e.g.,][]{tsai09}.

\subsection{Non-uniform distribution}
\begin{figure}
\centering
\includegraphics*[width=\linewidth]{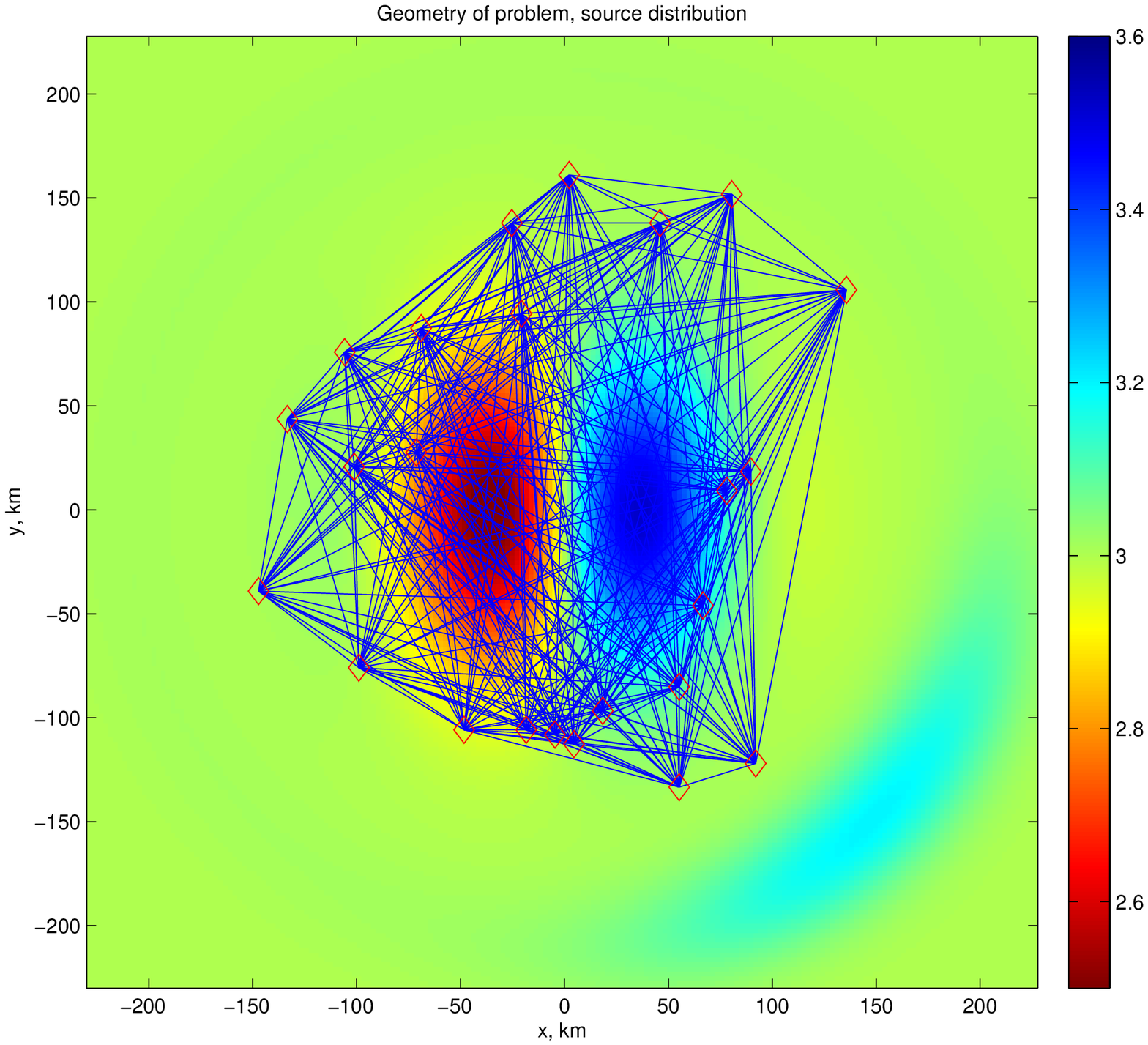}
\caption{The true source model anisotropically illuminates the network (from the South-East).
Stations are marked by diamonds and the wavespeed anomaly is in the center.
\label{source_anisotropic}}
\end{figure}

Figure~\ref{source_anisotropic} shows a non-uniform configuration of sources illuminating a network of 24 stations
which in turn surround the wavespeed anomaly. 

\begin{figure}
\centering
\includegraphics*[width=\linewidth]{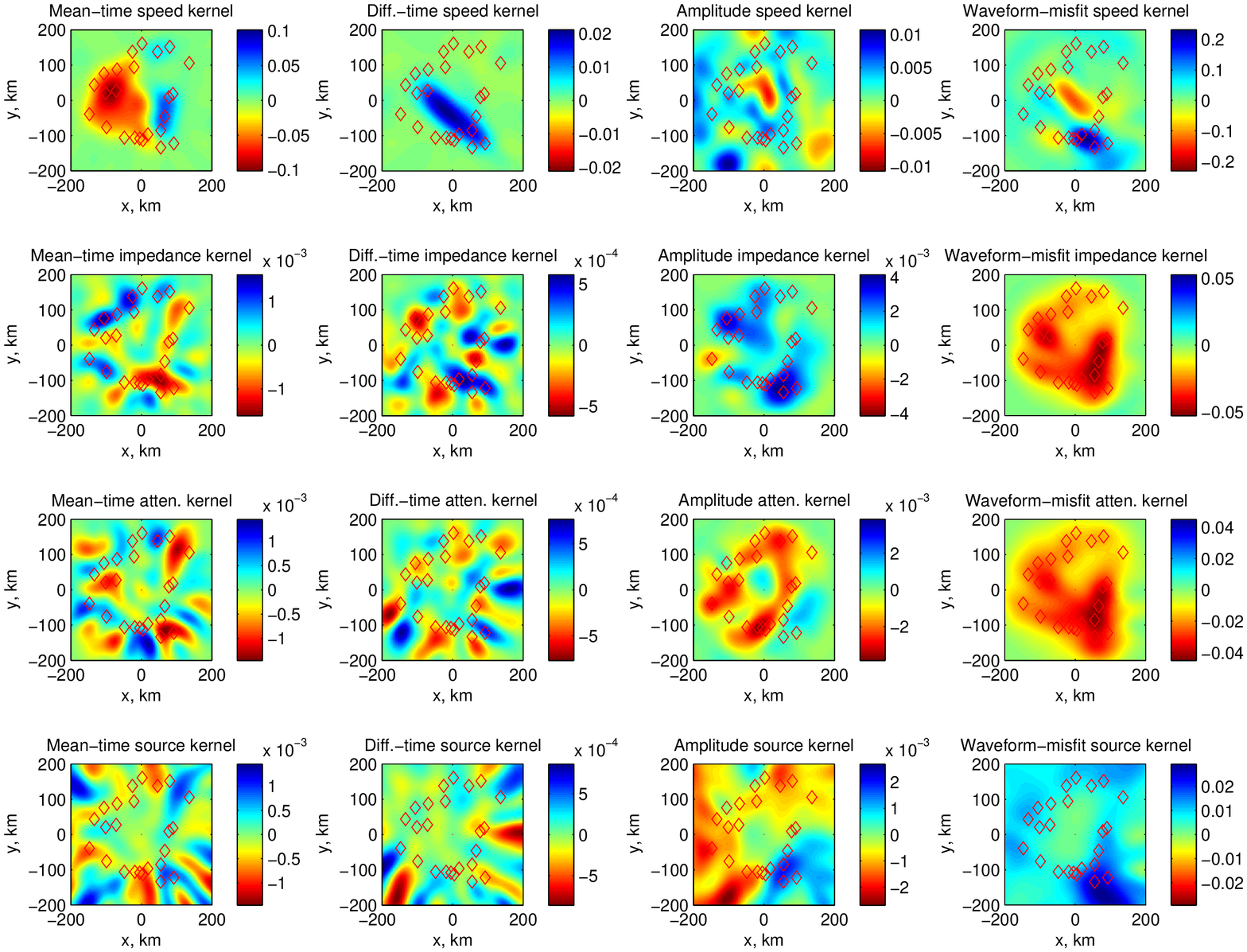}
\caption{Total gradients of misfit functions comprising different measurements with respect to model parameters. The true anomaly is in wavespeed.
The true source distribution is as shown in Figure ~\ref{source_anisotropic} whereas the starting source
model is as in Figure~\ref{source_uniform}.
The sum of all event kernels, in some ways, is the inversion, since it is the gradient of the entire misfit functional. 
There are four
parameters that we have imperfect knowledge of and that we wish to invert for: wavespeed, impedance, attenuation,
and source distribution (rows, top to bottom). We have four different measurements, mean and difference travel times, energies and waveform misfit (columns, left to right).
The true model contains positive and negative wavespeed perturbations to the tune of $\pm 20\%$ within the central region.
Mean travel times are
able to recover the wavespeed structure while the amplitude wavespeed kernels fail, owing to the fact that the starting source model isotropic whereas
the true source model is anisotropic. The difference travel-time gradient is aligned along North-West -- South-East direction, indeed suggesting
that anisotropic source distributions can create differences in travel times between the positive and negative branches \citep[e.g.,][]{hanasoge12_sources}.
The source-amplitude kernels suggest an increase in source magnitude in the South-East direction, which is what we would expect. Waveform differences, as before, are not to be trusted. A 3-point Gaussian smoothing filter was applied to the kernels.
\label{wrong_sources}}
\end{figure}

The set of total gradients for case (iii) are shown in Figure~\ref{wrong_sources}. Once again,
mean travel times are successful at imaging the wavespeed anomaly, although not to the
same degree as when the illumination was uniform, i.e., in Figure~\ref{event_wavespeed}.
The cross-correlation amplitude measurements suggest a deficit in source magnitude in all directions except the South-East
where it points to an increment.

Based on these simple tests, we form the following strategy
\begin{itemize}
\item Use cross-correlation energies to infer the azimuthal dependence of the source distribution (if not imaging its exact location),
\item This will serve as the starting model for the source distribution,
\item With this source-distribution, recompute kernels, 
\item With these kernels and mean travel times, invert for the wavespeed,
\item Possibly, after both of these steps, invert for attenuation/density.
\end{itemize}

\section{Wavespeed Inversions}
For all the problems, the starting structure model is a homogeneous uniform medium. We use a Gauss-Newton method to precondition the total gradient.
A one-step inversion is performed; see appendix~\ref{inverse.app} for details on the procedure. The model variance
is computed using the Hessian matrix, which we have full access to.

The experiments of section~\ref{expts1} emphasize the well known, namely that density and attenuation are indeed very difficult to invert for when the exact
source model is poorly known. In general however, the sensitivity to attenuation is relatively weak, making it a difficult quantity to measure. 
In this section we consider inverting only for source distribution followed by wavespeed and show that the correct theory can substantially
improve the recovery of anomalies. We study two more cases here 
\begin{enumerate}
\setcounter{enumi}{3}
\item Only wavespeed perturbation, true source distribution is nonuniform (as in Figure~\ref{source_anisotropic}),
comparison between source-wavespeed and only wavespeed inversions
\item Only wavespeed perturbation, true source distribution is nonuniform (as in Figure~\ref{source_anisotropic}), comparison
between cross-correlation and classical inversions
\end{enumerate}

\begin{figure}
\centering
\includegraphics*[width=\linewidth]{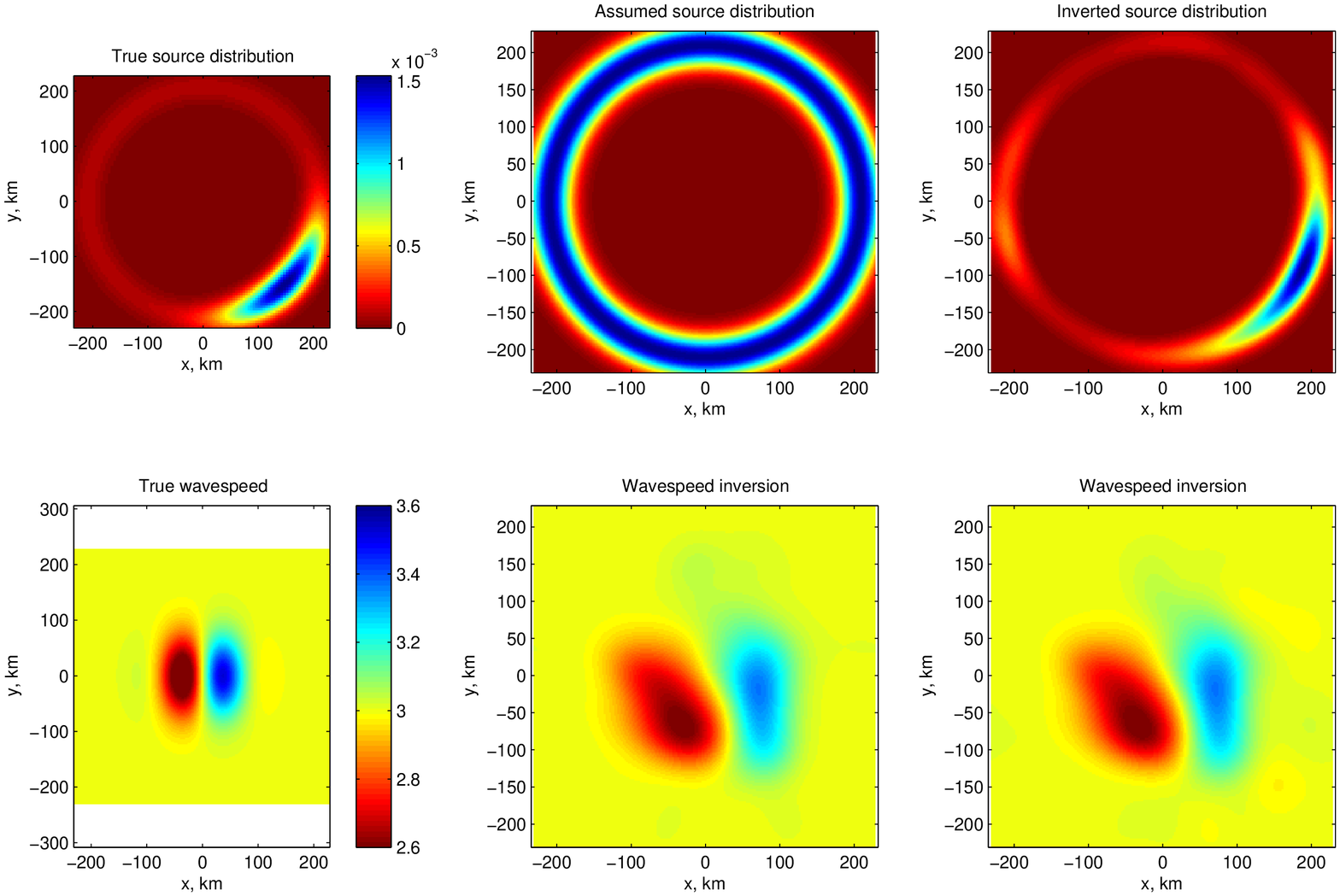}
\caption{Case (iv): a one-step Gauss-Newton inversion for the wavespeed. The color scale is the same
for all the plots on the bottom row.
The network, wavespeed anomaly and non-uniform source illumination are as in Figure~\ref{source_anisotropic}.
 The incomplete recovery of the anomaly is attributed to the sparsity of the network and the anisotropy of illumination.
In the middle column, we invert only for the wavespeed, assuming the source distribution is uniform. 
On the column on the right, we follow the strategy described at the end of section~\ref{expts1}, namely first
invert for the source distribution, followed by the wavespeed. The mean travel-time misfit, stated in
equation~(\ref{mean.tt}) decreases from 10.60 to 2.54. The energy misfit, in equation~(\ref{misfit.ener}),
when we invert for the source distribution on the right column drops from 1.04 to 0.59. 
The source inversion shows a slightly different azimuthal dependence from the true distribution because of
network does not have uniform azimuthal coverage.
Thus, although
the wavespeed models with and without a source inversion appear to be approximately the same, one 
can certainly aim to reduce the overall misfit of the system substantially by following this strategy.
We used 141 mean travel-time and 282 energy measurements in this inversion.
\label{correct_sources_inverse}}
\end{figure}

In case (iv), we study the impact of the source distribution on the reducing the misfits.
It is well known that a uniform ring of sources placed far away from a network mimics a general uniform source distribution
\citep[e.g.,][]{larose04}. Based on this argument, we our strategy for inverting for the source distribution is the following
\begin{itemize}
\item For most networks, the noise sources are too far away to image accurately \citep[e.g.,][]{hanasoge12_sources},
\item We therefore seek to reduce the dimensionality of the problem, parametrizing the sources as a ring with some azimuthal modulation in magnitude,
\item Construct a uniform ring of as large a radius as feasible surrounding the network,
\item Estimate the cross-correlation energy as a function of azimuth, interpolate, smooth and multiply the uniform ring by this function \citep[e.g.,][]{stehly06},
\item Recompute the synthetics, perform a source inversion and retain only the azimuthal information, multiplying the uniform ring
by this new azimuthal dependence.
\end{itemize}
Figure~\ref{correct_sources_inverse}
compares two cases with and without source inversions prior to the wavespeed inversion. The difference, in this particular
configuration, appears to not be significant, although the total misfit of the system, comprising energies and mean travel-times, fall
significantly where source anisotropies are accounted for. The incomplete recovery of the wavespeed anomaly may be attributed to 
the sparsity of the network and the anisotropy of illumination. Similarly, the azimuthal dependence of the source inversion is
different from that of the true distribution, owing to the fact that the network does cover all azimuths equally well.

\begin{figure}
\centering
\includegraphics*[width=\linewidth]{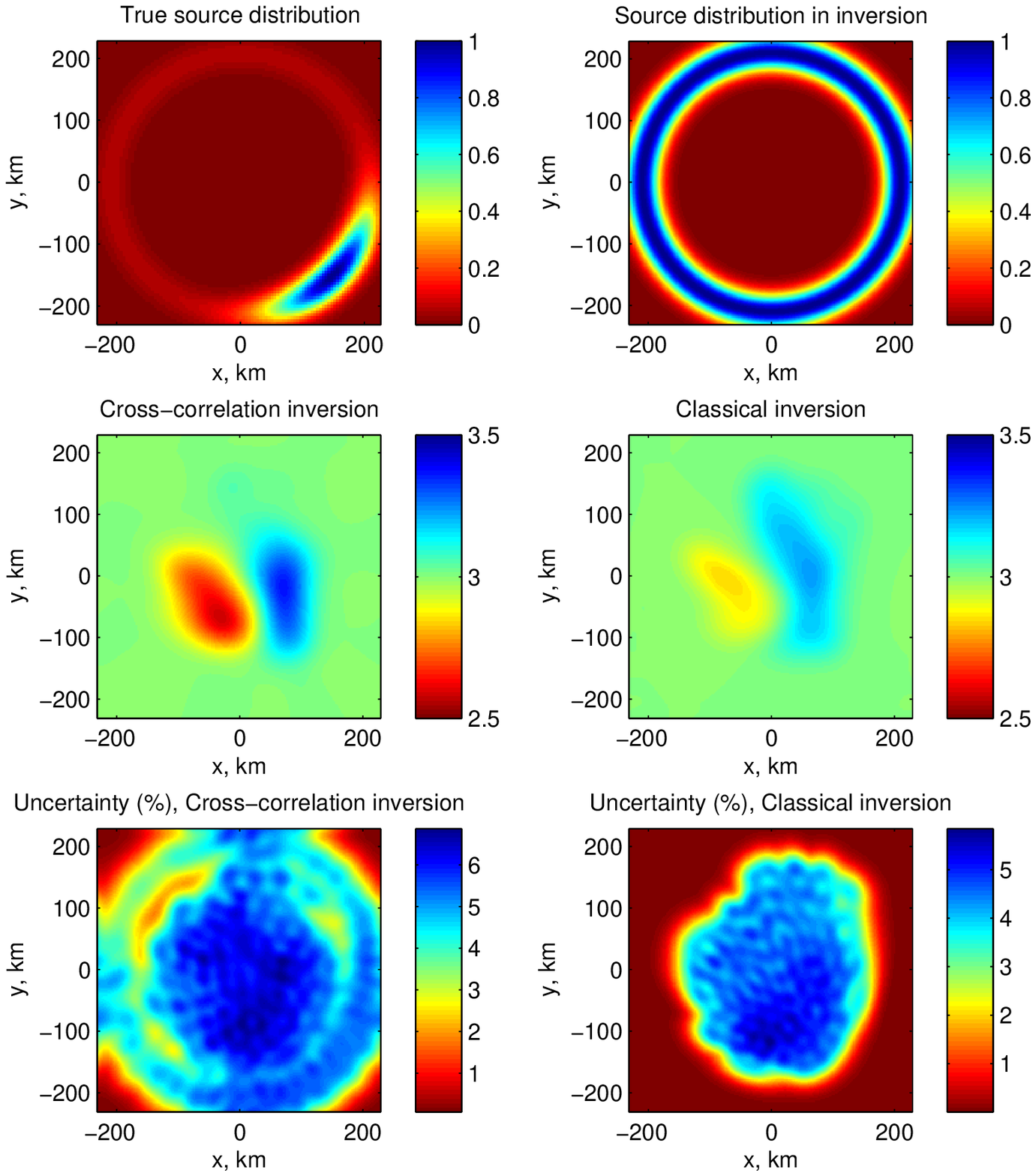}\vspace{-1.5cm}
\caption{Case (v): a one-step Gauss-Newton inversion for the wavespeed using 141 measurements of
mean travel times, comparison between correlation and classical approaches. 
The network, wavespeed anomaly and non-uniform source illumination are as in Figure~\ref{source_anisotropic}.
whereas the starting assumed source model is uniform, i.e., as in Figure~\ref{source_uniform}. 
The true model contains positive and negative wavespeed perturbations to the tune of $\pm 20\%$ within the central region.
We invert {\it only} for wavespeed using the method described in appendix~\ref{inverse.app} using correlation and classical approaches.
For the correlation approach, we compute cross-correlation kernels and for the classical approach, we treat the measurement
as if it were a travel time derived from raw wavefield measurements. The error in the theoretical model in the classical inversion propagates 
into the inversion and the misfit is larger. Applying the correct theoretical description is therefore quite important.
The Cram\'{e}r-Rao lower bound (appendix~\ref{inverse.app}) is shown in the lower-most panels for both inversions. The uncertainty
in inferring $\delta c/c$ is underestimated when using classical theory to interpret cross correlation measurements.
\label{classical_correlation}}
\end{figure}

Finally, in Figure~\ref{classical_correlation}, we compare the classical and correlation interpretations
and find the classical approach to underperform the correlation approach significantly. Indeed,
the correct theoretical description outweighs other aspects such as inverting for the source distribution
first.

\section{Conclusions}
Through the construction of a 2-D inverse problem, we have attempted to characterize
the sensitivity of cross-correlation measurements to parameters such as wavespeed,
density and attenuation. Using a Gauss-Newton method, we perform inversions for source
and wavespeed and demonstrate the utility of using cross-correlation theory as opposed
to applying classical interpretations to the measurements.
We summarize our conclusions here
\begin{itemize}
\item Density and attenuation are very difficult to measure, being sensitive to the source and structure model of choice,
\item The apertures of most small networks are insufficient to image sources that are typically very far away. We describe
a parametrization and method to address this issue while still accounting for anisotropic illumination,
\item We recommend inverting for sources using cross-correlation energies followed by using mean travel times of waves
to invert for the wavespeed. Performed in this order, we saw a factor of 2 reduction in amplitude misfit and a factor of 4
reduction in travel-time misfit,
\item Most of all, we found that a significant source of bias comes from not interpreting measurements correctly, i.e., treating
cross-correlations as if they were classical measurements,
\item A dominant source of uncertainty in the inversion is the sparsity of the station network, 
especially when using well defined arrivals for which we have a good starting model. We compute the 
uncertainty of the inversion using the diagonal of the approximate Hessian (see section~\ref{inverse.app}). This estimate
of the uncertainty may be considered a lower bound to the full uncertainty.
\end{itemize}

\section*{Acknowledgements}
S. M. H. is funded by NASA grant NNX11AB63G and would like to thank Courant Institute for their hospitality, J. Trampert
for some comments on a very early version of this manuscript that helped improve its scientific message. 

\bibliography{}
\bibliographystyle{gji}

\appendix

\section{Fourier Convention}\label{convention}
The following temporal Fourier transform convention is utilized
\begin{eqnarray}
\int_{-\infty}^\infty dt~e^{i\omega t}~ g(t) &=& {\hat g}(\omega) ,\\
\int_{-\infty}^\infty dt~e^{i\omega t} &=& 2\pi~\delta(\omega),\label{delta.func}\\
\frac{1}{2\pi}\int_{-\infty}^{\infty} d\omega~e^{-i\omega t}~ {\hat g}(\omega) &=& g(t),\label{inv.fourier}
\end{eqnarray}
where $g(t), {\hat g}(\omega)$ are a Fourier-transform pair. 
Using equation~(\ref{delta.func}), the equivalence between the cross-correlation in the temporal and a conjugate product in the Fourier domain is written so
\begin{equation}
h(t) = \int_{-\infty}^{\infty} dt'~ f(t')~ g(t+t') \Longleftrightarrow {\hat h}(\omega) = {\hat f}^*(\omega)~{\hat g}(\omega).\label{cross.c}
\end{equation}
The following relationship also holds (for real functions $f(t), g(t)$)
\begin{equation}
\int_{-\infty}^\infty dt~f(t)~g(t) = \frac{1}{2\pi}\int_{-\infty}^{\infty} d\omega~{\hat f}^*(\omega)~{\hat g}(\omega) = \frac{1}{2\pi}\int_{-\infty}^{\infty} d\omega~{\hat f}(\omega)~{\hat g}^*(\omega).\label{time.integral}
\end{equation}

In the spatial domain, we apply a similar convention
\begin{eqnarray}
\int_{\mathcal R^2} d\bx~e^{i{\bf k}\cdot\bx}~ g(\bx) &=& {\hat g}({\bf k}) ,\\
\int_{\mathcal R^2} d\bx~e^{i{\bf k}\cdot\bx} &=& (2\pi)^2~\delta({\bf k}),\label{delta.func.space}\\
\frac{1}{(2\pi)^2}\int_{\mathcal R^2} d{\bf k}~e^{-i{\bf k}\cdot\bx}\, {\hat g}({\bf k}) &=& g(\bx),\label{inv.fourier2}
\end{eqnarray}
where ${\bf k}$ is the 2-D wave vector, ${\hat g}({\bf k})$ and $g(\bx)$ are a Fourier-transform pair. The integral 
is over all 2-D space, i.e., ${\mathcal R^2} = [-\infty, \infty]\times[-\infty, \infty]$.
Using equation~(\ref{delta.func.space}), the equivalence between a convolution in the spatial and a product in Fourier domain are written thus
\begin{equation}
h(\bx) = \int_{\mathcal R^2}\, f(\bx')\, g(\bx-\bx') \Longleftrightarrow {\hat h}({\bf k}) =  {\hat f}({\bf k})~{\hat g}({\bf k}).\label{spaceconv}
\end{equation}

\section{Definitions of measurements and their variation}\label{varmisfit}
The definition of the travel-time \cite[e.g.,][]{gizon02, hanasoge11} is
\begin{equation}
\tau^\pm = \mp\frac{\int dt\,w^\pm(t)\,{\dot\ccref}\,(\cc - \ccref)}{\int dt\,w^\pm(t){\dot\ccref}^2},\label{tt.def}
\end{equation}
where $\tau^\pm$ is a travel-time shift associated with the positive or negative branch of the cross correlation respectively,
 $\ccref(t)$ and $\cc(t)$ are the reference and observed cross correlations for a station pair respectively, $w^\pm(t)$ is a windowing function applied to 
the positive or negative branches respectively, targeting
the travel-time shift of a specific arrival, and the over-dot notation indicates a temporal derivative, i.e., ${\dot\ccref} = \partial_t\ccref$. 

\begin{figure}
\centering
\includegraphics*[width=\linewidth]{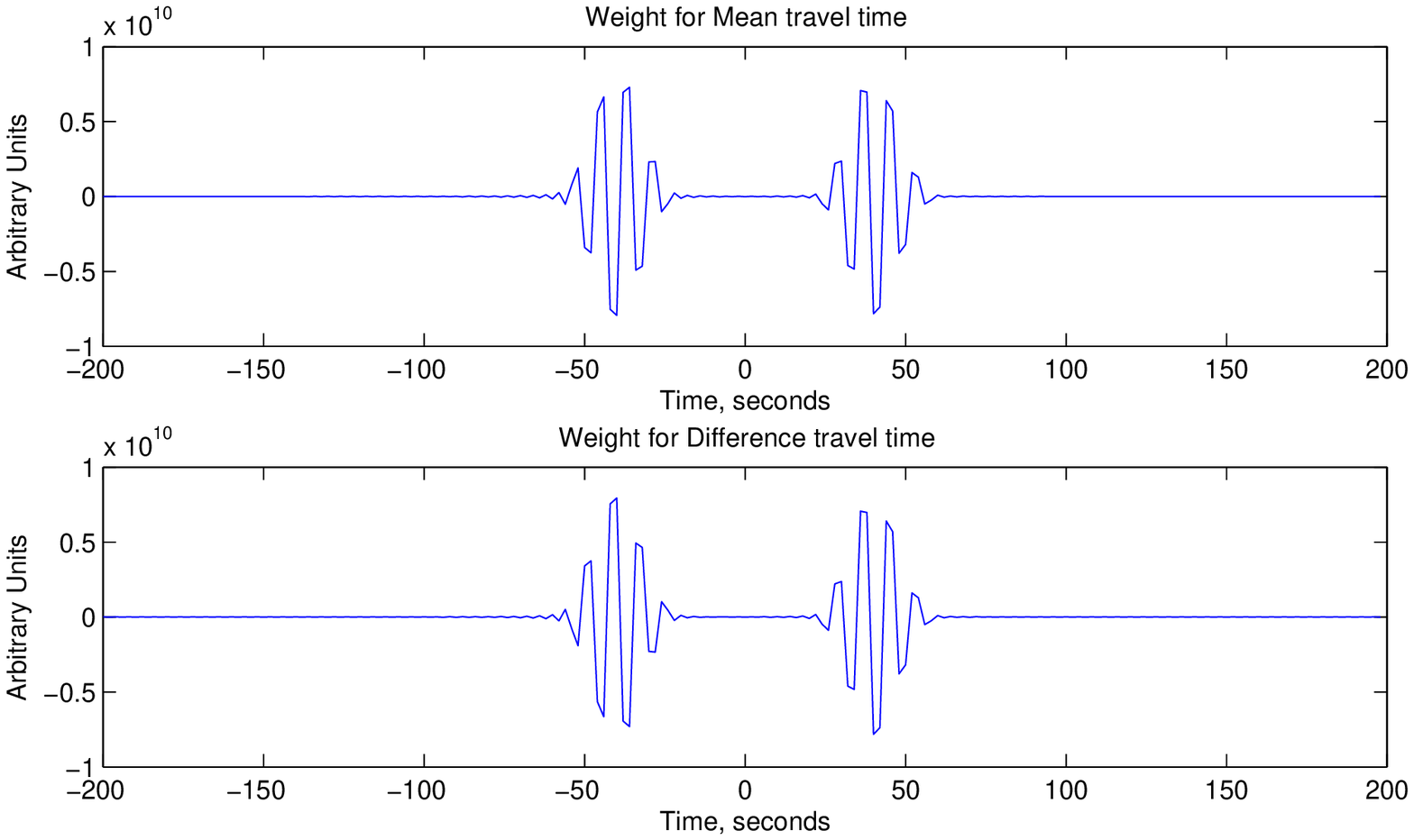}
\caption{Example weight function for mean and difference travel times for a station pair located 120 km apart (same configuration as in Figure~\ref{exam.cc}).
This weight function is used in computing mean and difference travel-time kernels and may be likened to the `adjoint source' as discussed by, e.g., \citet{tromp10}.
\label{weights_tt}}
\end{figure}

Travel times, energies and waveforms are the measurements used to compute kernels here. The travel time is calculated
using the definitions provided by, e.g., \citet{luo91, woodard,gizon02,tromp10,hanasoge11}. We recall the definition of 
the travel-time~(\ref{tt.def}). The variation of this quantity is given in equation~(\ref{tau.del}), and this allows us to derive the
weight function
\begin{equation}
{\mathcal W}(t) =  \mp\frac{w^\pm(t)\,{\dot\ccref}}{\int dt\,w^\pm(t){\dot\ccref}^2},
\end{equation}
giving us expressions for weight functions for the mean and difference travel times
\begin{eqnarray}
{\mathcal W}^m(t) &=&  \frac{1}{2}\left(\frac{w^+(t)\,{\dot\ccref}}{\int dt\,w^+(t){\dot\ccref}^2} + \frac{w^-(t)\,{\dot\ccref}}{\int dt\,w^-(t){\dot\ccref}^2}\right),\\
{\mathcal W}^d(t) &=&  \frac{1}{2}\left(\frac{w^+(t)\,{\dot\ccref}}{\int dt\,w^+(t){\dot\ccref}^2} - \frac{w^-(t)\,{\dot\ccref}}{\int dt\,w^-(t){\dot\ccref}^2}\right),
\end{eqnarray}
where $w^+(t)$ and $w^-(t)$ are windowing functions applied to isolate the relevant parts of the positive and negative branches respectively.
In Figure~\ref{weights_tt}, we show an example of weight functions for a station pair located 120 km apart.

The cross correlation energy is a non-linear quantity since it involves computing a non-linear function of the cross correlation. Since we are however
dealing with a linear problem, we linearize the variation of the functional along the lines of, e.g., \citet{dahlen02}, \citet{hanasoge12_sources}. 
Denoting the measured and reference cross correlation energies by $\ener$ and $\ener_{\rm ref}$, the amplitude misfit is given by
\begin{equation}
\chi= \frac{1}{2} \left(\ln\frac{\ener}{\ener_{\rm ref}}\right)^2,\label{misf}
\end{equation}
where we recall that the energy is defined as
\begin{equation}
\ener_{\rm ref} = \sqrt{\frac{\int dt\,\,w(t)\, {\cc_{\rm ref}(t)}^2}{\int dt\,w(t)} } = \sqrt{\frac{1}{2\pi T}\int d\omega\,{\cc}_{\rm ref}^*\, {\cc}_{\rm ref} },\label{syn.amp}
\end{equation}
where $w(t)$ is the windowing function, ${\mathcal C}$ is
the windowed cross correlation and $T = \int dt\,w(t)$.
To preserve simplicity, we do not apply frequency filters, although they may be easily included. 
With a little manipulation, the variation in misfit is given by
\begin{equation}
\delta\chi= -\left(\ln\frac{\ener}{\ener_{\rm ref}}\right) \frac{\delta \ener_{\rm ref}}{\ener_{\rm ref}} 
= -\left(\frac{1}{\ener_{\rm ref}}\right)^2 \left(\ln\frac{\ener}{\ener_{\rm ref}}\right)\frac{1}{2\pi T}\int d\omega\,{\cc}_{\rm ref}^* \, \delta{\cc}.
\end{equation}
Thus we define the energy weight function as 
\begin{equation}
{\mathcal W} = \frac{1}{T}\,\left(\frac{1}{\ener_{\rm ref}}\right)^2{{\cc}}_{\rm ref}.\label{weight}
\end{equation}

The waveform difference weight function is the simplest of the three, being equal to the windowed difference between the measured and reference
cross correlations.

\section{Expressions for Kernels and Validation}\label{kern.comp}
The wave operator ${\mathcal L}$ is given by
\begin{equation}
{\mathcal L} = -\omega^2\rho -i\omega\rho -\bnabla\cdot(\rho\,c^2\,\bnabla),
\end{equation}
and Green's function satisfies
\begin{equation}
{\mathcal L}G(\bx,\bx';\omega) = \delta(\bx-\bx'),
\end{equation}
where $\delta(\bx-\bx')$ is the delta function centered around $\bx = \bx'$.
The variation of Green's function as described in the single-scattering regime of the first Born approximation is given by
\begin{equation}
{\delta\mathcal L} G + {\mathcal L} \delta G = 0,
\end{equation}
and we have therefore
\begin{equation}
{\mathcal L} \delta G = -{\delta\mathcal L} G,
\end{equation}
and upon using Green's theorem, 
\begin{equation}
\delta G(\bx,\bx') = -\int d\bx{''}\,G(\bx,\bx{''})\,{\delta\mathcal L} G (\bx{''},\bx{'}).
\end{equation}
Let us first consider perturbations to the bulk modulus $\rho c^2$, i.e., $\delta{\mathcal L} = -\bnabla\cdot(\delta(\rho c^2)\bnabla)$,
where the spatial gradients and the bulk modulus are functions $\bx{''}$.
Recalling equations~(\ref{tau.del}) and~(\ref{struc.del}) and invoking identity~(\ref{time.integral}),
\begin{equation}
\delta\tau_{ij} = \int d\omega\,{\mathcal W}_{ij}^*\int d\bx'\,[\delta G^*(\bx_i, \bx')\, G(\bx_j, \bx') + G^*(\bx_i, \bx')\, \delta G(\bx_j, \bx')]\,S(\bx')\pspec.
\end{equation}
For simplicity, we consider one of the terms in the variation, with the analysis described here equally applicable to the other,
\begin{equation}
\int d\omega\,{\mathcal W}_{ij}^*\int d\bx' G^*(\bx_i, \bx')\, \delta G(\bx_j, \bx')\,S(\bx')\pspec = 
\int d\omega\,{\mathcal W}_{ij}^*\int d\bx' G^*(\bx_i, \bx')\,  S(\bx')\pspec\, \int d\bx G(\bx_j, \bx)\,\bnabla\cdot[\delta(\rho c^2)\bnabla G(\bx,\bx')].
\end{equation}
The inner integral over $\bx$ may be integrated by parts and using the fact that the medium is unbounded, boundary terms are dropped. This
allows us to free $\delta(\rho c^2)(\bx)$ from within the derivative. The result of this operation provides
\begin{equation}
\int d\omega\,{\mathcal W}_{ij}^*\int d\bx\,\delta(\rho c^2)\, \int d\bx' G^*(\bx_i, \bx')\, S(\bx')\,\pspec\, \bnabla G(\bx_j, \bx)\cdot\bnabla G(\bx,\bx'),
\end{equation}
which is rewritten as
\begin{equation}
\int d\bx\,\delta\ln(\rho c^2)\left\{  \int d\omega\,{\mathcal W}_{ij}^*\,\pspec\,\rho c^2\,\bnabla G(\bx_j, \bx)\cdot \int d\bx' G^*(\bx_i, \bx')\, \bnabla G(\bx,\bx')S(\bx')\right\}.
\label{temp.eq}
\end{equation}
Invoking seismic reciprocity and the assumption of a uniform starting background model, we recast equation~(\ref{temp.eq})
\begin{equation}
\int d\bx\,\delta\ln(\rho c^2)\left\{  \int d\omega\,{\mathcal W}_{ij}^*\,\pspec\,\rho c^2\,\bnabla G(\bx,\bx_j)\cdot \int d\bx' S(\bx') G^*(\bx', \bx_i)\, \bnabla G(|\bx-\bx'|)\right\}
\end{equation}
where the term within the parentheses contributes to the bulk modulus kernel.
The inner integral over $\bx'$ is the crux of the scattering problem, and represents the primary challenge in computing these kernels. Because we have assumed translation
invariance, we may rewrite this in spatial Fourier domain, converting the convolution to a Fourier product (see Appendix~\ref{convention}, {specifically
relation~[\ref{spaceconv}]}) and transforming back to the spatial domain. 
We apply similar analyses to the density and attenuation terms. The expressions for the three kernels are as follow
\begin{eqnarray}
K_{\rho c^2}(\bx) &=& \rho c^2\,\int d\omega\,{\mathcal W}_{ij}^*\,\pspec\,\left[\bnabla G(\bx,\bx_j)\cdot \int d\bx' S(\bx')\, G^*(\bx', \bx_i)\, \bnabla G(|\bx-\bx'|)+\right. \nonumber\\
&&\left.\bnabla G^*(\bx,\bx_i)\cdot \int d\bx' S(\bx')\, G(\bx', \bx_j)\, \bnabla G^*(|\bx-\bx'|) \right]\label{kern.bulk},\\
K_{\rho\Gamma}(\bx) &=& -\rho \Gamma\,\int d\omega\,i\omega\,{\mathcal W}_{ij}^*\,\pspec\,\left[ G(\bx,\bx_j) \int d\bx' S(\bx')\, G^*(\bx', \bx_i)\, G(|\bx-\bx'|)-\right. \nonumber\\
&&\left. G^*(\bx{''},\bx_i) \int d\bx' S(\bx')\, G(\bx', \bx_j)\, G^*(|\bx-\bx'|) \right]\label{kern.atten},\\
K_{\rho}(\bx) &=& -\rho \,\int d\omega\,\omega^2\,{\mathcal W}_{ij}^*\,\pspec\,\left[ G(\bx,\bx_j) \int d\bx' S(\bx')\, G^*(\bx', \bx_i)\, G(|\bx-\bx'|)+\right. \nonumber\\
&&\left. G^*(\bx,\bx_i) \int d\bx' S(\bx')\, G(\bx', \bx_j)\, G^*(|\bx-\bx'|) \right]\label{kern.dens}.
\end{eqnarray}
All put together, we recover the following relation between travel-time shifts and the kernels
\begin{equation}
\delta\tau = \int d\bx\, \delta\ln(\rho c^2) K_{\rho c^2} + \delta\ln(\rho\Gamma)\, K_{\rho\Gamma} + \delta\ln\rho\, K_\rho,
\end{equation}
where $K_{\rho c^2}$ is the bulk modulus kernel, $K_{\rho \Gamma}$ is a density weighted attenuation kernel and $K_\rho$ is the density kernel.
These are not necessarily the best choices for parameters to invert for \citep[e.g.,][]{zhu09}. Consequently, we rearrange the terms thus
\begin{equation}
\delta\tau = \int d\bx\,  2 K_{\rho c^2}\,\delta\ln c +  K_{\rho\Gamma}\,\delta\ln\Gamma  + (K_{\rho c^2} + K_{\rho\Gamma} + K_\rho)\delta\ln\rho,
\end{equation}
and define the following as the wavespeed, attenuation and impedance kernels respectively
\begin{eqnarray}
K_c &=& 2 K_{\rho c^2}\label{kern.c},\\
K_\Gamma &=& K_{\rho \Gamma},\label{kern.gam}\\
K'_\rho &=& K_{\rho c^2} + K_{\rho\Gamma} + K_\rho.\label{kern.imp}
\end{eqnarray}
The simplest kernel of all is the source kernel \cite[see, e.g.,][]{hanasoge12_sources}, which has no scattering terms
\begin{equation}
K_{\rm S}(\bx) = \int d\omega\,{\mathcal W}_{ij}^*\,\pspec\, G^*(\bx, \bx_i)\,G(\bx, \bx_j).\label{source.kernel}
\end{equation}
All of these kernels put together, the variation of the travel time reduces to
\begin{equation}
\delta\tau = \int d\bx \,  K_{c}\,\delta\ln c +  K_{\Gamma}\,\delta\ln\Gamma  + K'_{\rho}\,\delta\ln\rho + K_{\rm S}\, \delta{\rm S}.
\end{equation}
In order to validate travel-time kernels, we consider a fractional uniform change to the wavespeed, e.g., $\delta\ln c = 0.001$. The measured change in the travel time 
must then be equal to the integral of the wavespeed kernel, i.e.,
\begin{equation}
\delta\tau  = 0.001 \int d\bx\,K_{c}.
\end{equation}
Similarly, for the impedance kernel, a fractional uniform change in density can lead only to a correspondingly small change in the amplitude of the cross correlation, with
there being virtually no change in the travel time as defined in equation~(\ref{tt.def}).Consequently, the integral of the impedance kernel is virtually zero. We use both 
of these tests to verify the kernel computation procedure. A test for the source kernels computed for the energy measurement was described in \citet{hanasoge12_sources},
requiring
\begin{equation}
\int d\bx\,K_{\rm S}\, S = 1,
\end{equation}
where ${\rm S}$ is the source distribution.

\section{Gauss-Newton Method and Hessian-Derived Uncertainties}\label{inverse.app}

For simplicity's sake, let us rewrite the misfit $\chi$ as a sum of $M$ individual measurement misfits $\mu_k$, i.e., $\chi =  \sum_{i,j} \Lambda_{ij}\, \mu_i\,\mu_j/2 = {\bmu}^T\blamb\bmu/2$,
where $\bmu$ is an $M\times 1$ vector and $\blamb$ is an $M\times M$ sized inverse of the data-noise covariance matrix.
Expanding the misfit functional, which is dependent on the properties of the medium $\bm(\bx)$, around small variations $\delta \bm({\bf x})$,
\begin{equation}
\chi(\bm + \delta \bm) = \chi(\bm)  + \frac{\partial\chi}{\partial m_k}\delta m_k + \frac{1}{2}\frac{\partial^2\chi}{\partial m_k \partial m_\ell}\delta m_k\delta m_\ell + O(||\delta f||^3),
\end{equation}
where $m_k = m(x_k)$ in which we have $N$ points on the grid ${\bf x}$ and we invoke Einstein's convention of summing over repeated indices. We denote the Jacobian $\frac{\partial \mu_i}{\partial m_k} = -K_i(x_k)$, where the right side is the kernel for measurement $i$ evaluated at location $j$.
The gradient of the misfit function with respect to model parameters is
\begin{equation}
\frac{\partial\bmu}{\partial{\bm}} = {\bf K},
\end{equation}
which is an $M\times N$ matrix and $f$ is an $N\times 1$ vector. The first term of the Taylor expansion when written in matrix notation is
\begin{equation}
\frac{\partial\chi}{\partial \bm}\delta \bm =\delta \bm^T\,{\bf K}^T\,\blamb\,\bmu,
\end{equation}
is a $1\times N$ vector.
The second order term is
\begin{equation}
\frac{\partial^2\chi}{\partial \bm\,\partial \bm}\delta \bm\, \delta \bm \approx \delta \bm^T\,{\bf K}^T\,\blamb\,{\bf K}\,\delta\bm.
\end{equation}
Thus the Gauss-Newton update (upon seeking the stationary point of $\delta\chi$) becomes
\begin{equation}
{\bf K}^{\rm T}\blamb{\bf K}\,\delta \bm = - {\bf K}^{\rm T}\blamb\,\bmu.
\end{equation}
The covariance matrix is positive definite and so it has a real (symmetric) square root. Defining $\tilde{\bf K} = \sqrt{\blamb}\,{\bf K}$, we obtain
\begin{equation}
\tilde{\bf K}^{\rm T}\tilde{\bf K}\,\delta  \bm = - \tilde{\bf K}^{\rm T}\,\sqrt{\blamb}\,\bmu.
\end{equation}
The right side is the negative of the total gradient, or sum of all kernels. 
The left side may be solved using a sparse matrix inversion method. Consider $M$ measurements $\ll N$ spatial points. So the full kernel matrix is $M\times N$. Performing
a singular value decomposition $\{\tilde{\bf K}\} = U\Sigma V^{\rm T}$ and we have $\tilde{\bf K}^{\rm T}\tilde{\bf K} = V\Sigma^2V^{\rm T} =  \sum_i\sigma_i^2\,v_i\,v^{\rm T}_i$. 
We note that $v^{\rm T}_j v_k = \delta_{jk}$. Adding damping $d$ to the singular values 
we obtain the update 
\begin{equation}
\{\delta \bm\} = -\sum_i ({\sigma}_i + d)^{-2}\,v_i\,v_i^{\rm T} \tilde{\bf K}^{\rm T}\,\sqrt{\blamb}\,\bmu.
\end{equation}

The diagonal of inverse of the Hessian ${\rm diag}[({\bf K}^{\rm T}\,\blamb\,{\bf K})^{-1}]$ 
gives us the total model variance $\sigma_m^2$ as a function of space \citep[assuming multivariate Gaussian statistics, e.g.,][]{tarantola87},
\begin{equation}
\{\sigma_m^2\} = {\rm diag}\left[\sum_i ({\sigma}_i + d)^{-2}\,v_i\,v_i^{\rm T}\right].\label{modvar}
\end{equation}
The model variance provided by equation~(\ref{modvar}) is the Cram\'{e}r-Rao bound where the Fisher information matrix is the Hessian. Note
that the full model covariance is obtained by evaluating the inverse of the Hessian.

\end{document}